\documentclass[letterpaper, 10pt, conference]{ieeeconf}

\IEEEoverridecommandlockouts
\usepackage{amsmath,amssymb,amsfonts}
\usepackage{tikz}
\usepackage{float}
\usepackage{circuitikz}
\let\labelindent\relax 
\usepackage{enumitem}
\usetikzlibrary{arrows.meta,positioning,calc,fit}
\newtheorem{assumption}{Assumption}
\newtheorem{theorem}{Theorem}
\newtheorem{lemma}{Lemma}
\newtheorem{remark}{Remark}
\newcommand{\R}{\mathbb{R}}
\newcommand{\N}{\mathbb{N}}
\newcommand{\X}{\mathcal{X}}
\newcommand{\Pip}{\Pi_p}
\DeclareMathOperator{\tr}{tr}
\newcommand{\lmin}{\lambda_{\min}}

\title{\LARGE \bf
Robust State-Parameter Estimation in Second-Order Th\'evenin Batteries}
\author{Wissam Khennoune, G. Q. Bao Tran, Thach Ngoc Dinh, J\'er\'emy Van Gorp, and Mathieu Moze 
\thanks{W.~Khennoune, T.~N.~Dinh, J.~Van Gorp, and M.~Moze are with Conservatoire National des Arts et M\'etiers (CNAM), CEDRIC-Laetitia, 292 rue St-Martin, 75141 Paris, France  (e-mail: {\tt\{wissam.khennoune, ngoc-thach.dinh, jeremy.vangorp, mathieu.moze\}@lecnam.net}). G.~Q.~B.~Tran is with Coordinated Science Laboratory, University of Illinois Urbana-Champaign, Urbana, IL 61801, USA (e-mail: {\tt baotran@illinois.edu}). This research is supported in part by a fellowship by the Hanse-Wissenschaftskolleg--Institute for Advanced Study (HWK), and in part by the AFOSR MURI FA9550-23-1-0337 and NSF CMMI-2452534 grants.}}

\begin{document}

\maketitle
\thispagestyle{empty}
\pagestyle{empty}

\begin{abstract}
We propose a scheme for estimating the full states and parameters of a second-order Th\'evenin battery
model, using distinct features of the measured current. An
identification step, run during suitable rest intervals, identifies the resistances and capacitances
of the two RC branches
from the voltage relaxation by linear least
squares, without knowledge of the open-circuit voltage (OCV) map. An online
estimator then runs continuously: the
polarization voltages are estimated by open-loop observers; the series
resistance is updated at current jumps using least squares; the state of charge (SOC) is propagated by Coulomb counting
using the current capacity estimate, and both estimates are updated periodically by a
sliding-window regression. We show that the identification is exact under a mild rank condition, that the series resistance is recovered at the first current jump, and that the online state and capacity estimates converge exponentially under verifiable excitation conditions. We also establish
robustness of the online estimator under bounded process disturbances and measurement noise.
\end{abstract}

 \section{INTRODUCTION}
 \label{intro}
Lithium-ion batteries have become a leading energy-storage technology for electric vehicles thanks to their high energy density, low self-discharge, and long cycle life~\cite{Du2016}. Their safe, efficient operation requires a battery management system (BMS) to monitor state of charge (SOC), capacity, and internal resistances. However, these quantities cannot be measured directly and must be estimated, mainly from current and terminal voltage. Moreover, capacity and electrical parameters evolve with aging and operating conditions, making online estimation important for reliable monitoring.

SOC estimation approaches include Coulomb counting, open-circuit-voltage (OCV)-based methods~\cite{cui2022extended}, data-driven techniques~\cite{wu2025advancements,ofoegbu2025state}, and model-based methods~\cite{lahme2024interval,khennoune2026interval}. Coulomb counting is simple but requires initial SOC and battery capacity, and is sensitive to accumulated current-measurement errors. OCV-based methods use the voltage--SOC relationship, generally require relaxation toward equilibrium, and can lose accuracy in relatively flat OCV regions. Data-driven methods capture complex battery behavior but often require large, representative datasets and offline training~\cite{Lipu2020}. Model-based methods combine current and voltage measurements with mathematical battery dynamics. Equivalent-circuit models (ECMs) balance modeling accuracy and computational complexity~\cite{Nejad2016}. The second-order Th\'evenin model represents polarization dynamics over two time scales using two RC branches and only three states: SOC and two polarization voltages. It captures richer transients than a first-order model without higher-order complexity, making it suitable for BMS implementation~\cite{mohamed2025advancements,de2026linear}. ECMs commonly employ Kalman-based estimators, including extended and unscented Kalman filters~\cite{chen2012state,yu2017lithium,beelen2020joint}, and nonlinear observers. Sliding-mode observers have been investigated for SOC estimation under practical operating conditions~\cite{riva2026design}. At pack level,~\cite{khalil2026estimation} proposed a hybrid minimum/maximum cell SOC estimator with robustness guarantees under disturbances and measurement noise. These approaches highlight the importance of estimation accuracy, convergence, and robustness in practical BMS applications.

Accurate state estimation also depends on uncertain electrical parameters that vary with aging and operating conditions, motivating combined online parameter identification and SOC estimation. Work~\cite{ouyang2019state} combined robust recursive least-squares (RLS) identification with a robust SOC observer for a first-order Th\'evenin model, assuming known capacity and sufficiently informative current excitation. Paper~\cite{naseri2021enhanced} combined a Wiener-based second-order battery model with extended Kalman filter-based RLS for parameter identification and an EKF for SOC estimation, also assuming known capacity. Recently,~\cite{liu2025online} proposed decoupled RLS with an adaptive forgetting factor for online identification of the five second-order Th\'evenin electrical parameters, excluding capacity. Paper~\cite{desai2026adaptive} proposed an adaptive multi-rate moving-horizon estimator (MHE) for joint SOC and parameter estimation with a first-order Th\'evenin model. Wavelet-based sampling and parallel MHE structures reduce computation, while nominal capacity is assumed known.

Despite these advances, the cited works do not jointly estimate \emph{all} five second-order Th\'evenin electrical parameters, SOC, and capacity (see Section~\ref{sec:model}). RLS-based methods generally require persistent excitation of regression signals to guarantee parameter convergence, often difficult to assess from driving profiles. MHE requires online optimization, with convergence guarantees depending on its formulation and assumptions~\cite{wynn2014convergence}. For instance,~\cite{desai2026adaptive} uses a first-order model with known capacity. These approaches also do not exploit rest intervals, current jumps, and charge throughput as separate information sources for updating estimates. This motivates a hybrid scheme estimating each quantity when the corresponding current feature becomes available.

We propose a hybrid estimation scheme for a second-order Th\'evenin model that fully identifies electrical parameters and estimates SOC and battery capacity online. It strategically exploits three measured-current \emph{features}: rests, jumps, and charge throughput. During rests, linear least squares identifies the two RC resistances without the OCV map, or all four RC parameters when capacitances are unknown. Current jumps yield the series resistance, while Coulomb counting and periodic sliding-window regression of an inverted OCV signal jointly estimate SOC and capacity. We establish suitable, mild conditions, nominal exponential convergence, and robustness of the estimation under bounded disturbances and measurement noise. An EPA Urban Dynamometer Driving Schedule (UDDS) current profile~\cite{EPA_UDDS} illustrates the method. The reproducible MATLAB code is available at~\cite{code}.

 \emph{Notations.} Let $|\cdot|$ denote the Euclidean vector norm
and $\|\cdot\|$ the induced matrix norm. For a symmetric matrix $M$ (whose eigenvalues are real),
$\lmin(M)$ is its smallest eigenvalue, $\det M$ its determinant, and
$\tr M$ its trace.

\section{SYSTEM DESCRIPTION AND ANALYSIS}
\label{sec:model}

\subsection{Battery Model}
We consider the second-order Th\'evenin equivalent-circuit model shown in
Fig.~\ref{fig:thevenin_2rc}.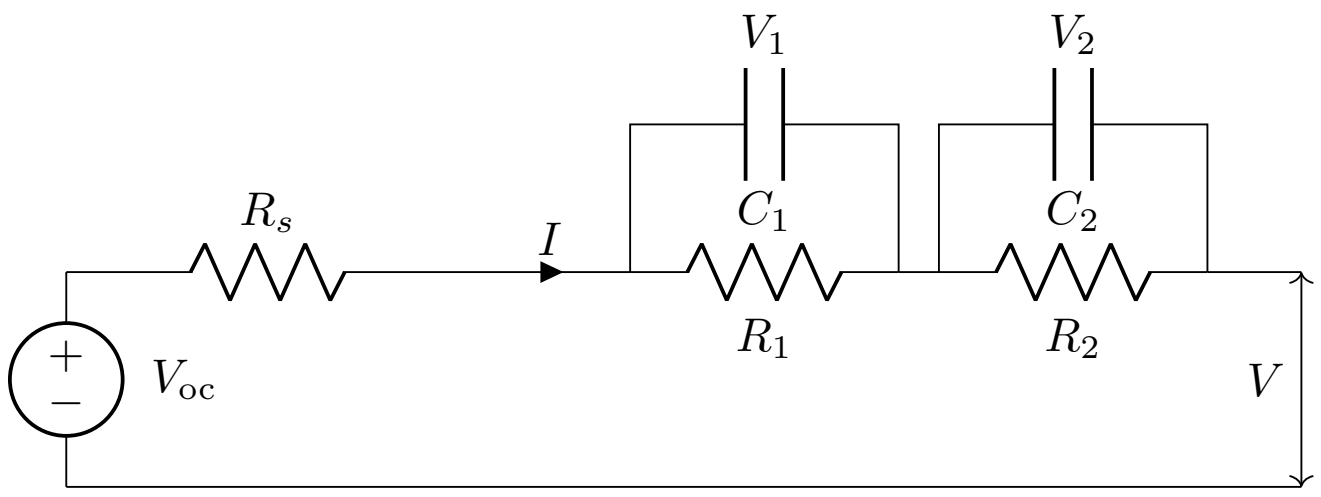
\begin{figure}[!ht]
    \centering
    \resizebox{\columnwidth}{!}{%
\begin{circuitikz}[american, font=\normalsize]
    \draw
    (0,1.6) to[vsource, v^=$V_{\mathrm{oc}}$] (0,0)
    (0,1.6) to[R=$R_s$] (3,1.6)
    to[short, i=$I$] (4.2,1.6);
    \draw
    (4.2,1.6) to[R, l_=$R_1$] (6.2,1.6);
    \draw
    (4.2,1.6) to[short] (4.2,2.7)
    to[C] (6.2,2.7)
    to[short] (6.2,1.6);
    \node at (5.2,2.05) {$C_1$};
    \node[anchor=south, inner sep=0pt] at (5.2,3.22) {$V_1$};
    \draw
    (6.2,1.6) to[short] (6.5,1.6);
    \draw
    (6.5,1.6) to[R, l_=$R_2$] (8.5,1.6);
    \draw
    (6.5,1.6) to[short] (6.5,2.7)
    to[C] (8.5,2.7)
    to[short] (8.5,1.6);
    \node at (7.5,2.05) {$C_2$};
    \node[anchor=south, inner sep=0pt] at (7.5,3.22) {$V_2$};
    \draw
    (8.5,1.6) to[short] (9.2,1.6);
    \draw[<->] (9.2,1.6) -- (9.2,0) node[midway,left] {$V$};
    \draw
    (9.2,0) to[short] (0,0);
\end{circuitikz}%
}
    \caption{Second-order Th\'evenin equivalent-circuit model.}
    \label{fig:thevenin_2rc}
\end{figure}

The SOC dynamics follow from Coulomb counting, while the dynamics of the two RC branches and the terminal voltage are obtained from Kirchhoff's laws. Defining the state $x:=(x_1,x_2,x_3)$, we obtain the model
\begin{subequations}\label{eq:battery}
\begin{align}
    \dot x_1 &= -\theta u,
    \label{eq:battery_x1}\\
    \dot x_2 &= -a x_2+g_1u,
    \label{eq:battery_x2}\\
    \dot x_3 &= -b x_3+g_2u,
    \label{eq:battery_x3}\\
    y &= p(x_1)-x_2-x_3-R_su.
    \label{eq:battery_output}
\end{align}
\end{subequations}
Here, $x_1$ is the state of charge (SOC), while $x_2$ and $x_3$
are the polarization voltages $V_1$ and $V_2$, respectively,
across the two RC branches in Fig.~\ref{fig:thevenin_2rc}.
The input is the measured current $u=I$ (positive during discharge),
and the output is the measured terminal voltage $y=V$. We let $\theta:=1/Q$, where
$Q>0$ is the battery capacity, and denote the series resistance by $R_s>0$. The RC
branch parameters enter the model through
$$
    a=1/(R_1C_1),\;\;
    b=1/(R_2C_2),\;\;
    g_1=1/C_1,\;\;
    g_2=1/C_2.
$$
For simulations, we use the parameters from~\cite{chen2012state}. They are the constants
$R_s=0.0013$~($\Omega$), $R_1=0.0042$~($\Omega$), $R_2=0.0020$~($\Omega$),
$C_1=17111$~(F), $C_2=440.5675$~(F), and
$Q=72000$~(C), i.e., $20$~(Ah). The open-circuit voltage (OCV) map $p$ is the
seventh-order polynomial
\begin{multline*}
p(x_1) =8.4073x_1^7 - 19.892x_1^6 + 11.497x_1^5
+ 4.161x_1^4 \\- 4.5533x_1^3 + 0.34365x_1^2
+ 0.64685x_1 + 3.5016 .
\end{multline*}
The time constants are $1/a=R_1C_1\approx72$~(s) and
$1/b=R_2C_2\approx0.88$~(s), so that $a<b$ and the first branch is the slower
one. We assume this ordering throughout to assign the identified time constants to
their respective branches. The map
$p$ is strictly increasing on $\X:=[0,1]$, and we
let $m_p:=\min_{z\in\X}p^\prime(z)>0$ with $p^\prime$ the derivative of $p$.

In this work, we propose a complete joint state-parameter estimation scheme. We identify
$R_1$ and $R_2$, as well as $C_1$ and $C_2$ if they are unknown, and estimate $R_s$,
$Q$, and the state $x(t)$ online. The \emph{only} model knowledge required is the OCV
map $p$, which is used by the online estimator but not by the identification. We make
the following assumptions.
 
\begin{assumption}
\label{ass:operation}
The current $u$ is bounded, i.e.,  $|u(t)|\leq\overline u$ for all $t\geq0$ for some constant $\overline u > 0$, and piecewise
continuous with finitely many discontinuities on every bounded
interval. The SOC satisfies $x_1(t)\in\X$ for all $t\geq0$.
\end{assumption}

\subsection{Useful Features of Current Profile}
\label{subsec:udds}

\begin{figure}[ht]
    \centering    \includegraphics[width=\columnwidth]{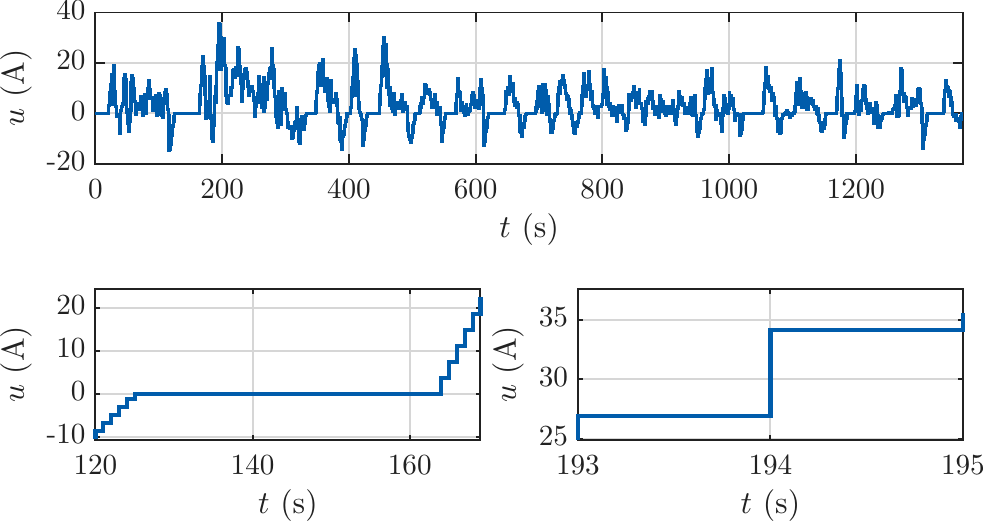}
    \caption{UDDS-derived current profile. Top: full driving cycle. Bottom: enlarged view of a rest with $u=0$ (left) and of a current jump (right)}
    \label{fig:udds_current}
\end{figure}

To examine the battery under urban driving conditions,
we consider the UDDS provided by~\cite{EPA_UDDS}. The prescribed speed
profile is converted into battery current using a
simplified vehicle model. The resulting
current $u$, shown in Fig.~\ref{fig:udds_current},
alternates between discharging, charging, and rest
periods.

In this UDDS profile of more than $23$ minutes, the current
contains more than $10$ zero-current (or rest) intervals lasting at least
10~(s), together with more than $900$ jumps exceeding $0.5$~(A)
in magnitude. During rests, the terminal voltage
relaxes as the polarization voltages decay, revealing
the dynamics of the two RC branches. Current jumps
produce abrupt changes in the ohmic voltage drop,
whereas the charge exchanged during operation
governs the SOC evolution through Coulomb counting.
These distinct operating conditions provide
complementary information about the battery states
and parameters. More precisely, the following observations are exploited for
estimation.
\begin{enumerate}[leftmargin=*,nosep]
\item \emph{Rest intervals} ($u\equiv0$). $x_1$ is constant and $x_2$, $x_3$
decay to $0$ with the mentioned time constants, so that $y$ is an unknown constant plus two
decaying exponentials that depend neither on $R_s$ nor on $Q$. The voltage
relaxation identifies $R_1$ and $R_2$ when the capacitances $C_1$, $C_2$ are known. If they are
unknown, two rests are used to identify all four
parameters (see Section~\ref{sec:rc_id});
\item \emph{Current jumps} ($u(t_k^+)\neq u(t_k^-)$). The current profile is specified
at $1$~(Hz), so that
$u$ is piecewise constant. Since the state $x$ is continuous, $y$ jumps
only through the ohmic drop, by $-R_s\delta u_k$, where $\delta
u_k=u(t_k^+)-u(t_k^-)$. This relation is used to identify $R_s$
(see Section~\ref{sec:online});
\item \emph{Charge throughput} (varying accumulated charge). By Coulomb
counting, $x_1$ is affine in
the accumulated charge with slope $-\theta$, so that a window over
which the charge varies carries information on $x_1$ and $Q$
(Section~\ref{sec:online}).
\end{enumerate}
Our scope covers current profiles other than UDDS, provided that they also have these features. Measurement errors are of two kinds. Errors entering the dynamics, such as
current-sensor bias and noise, affect the states, which remain
continuous. Voltage-sensor noise acts directly on $y$ and may
jump, in particular at the jump times of $u$
(see Section~\ref{sec:robust}).
 
\section{ESTIMATION SCHEME}
\label{sec:observer}
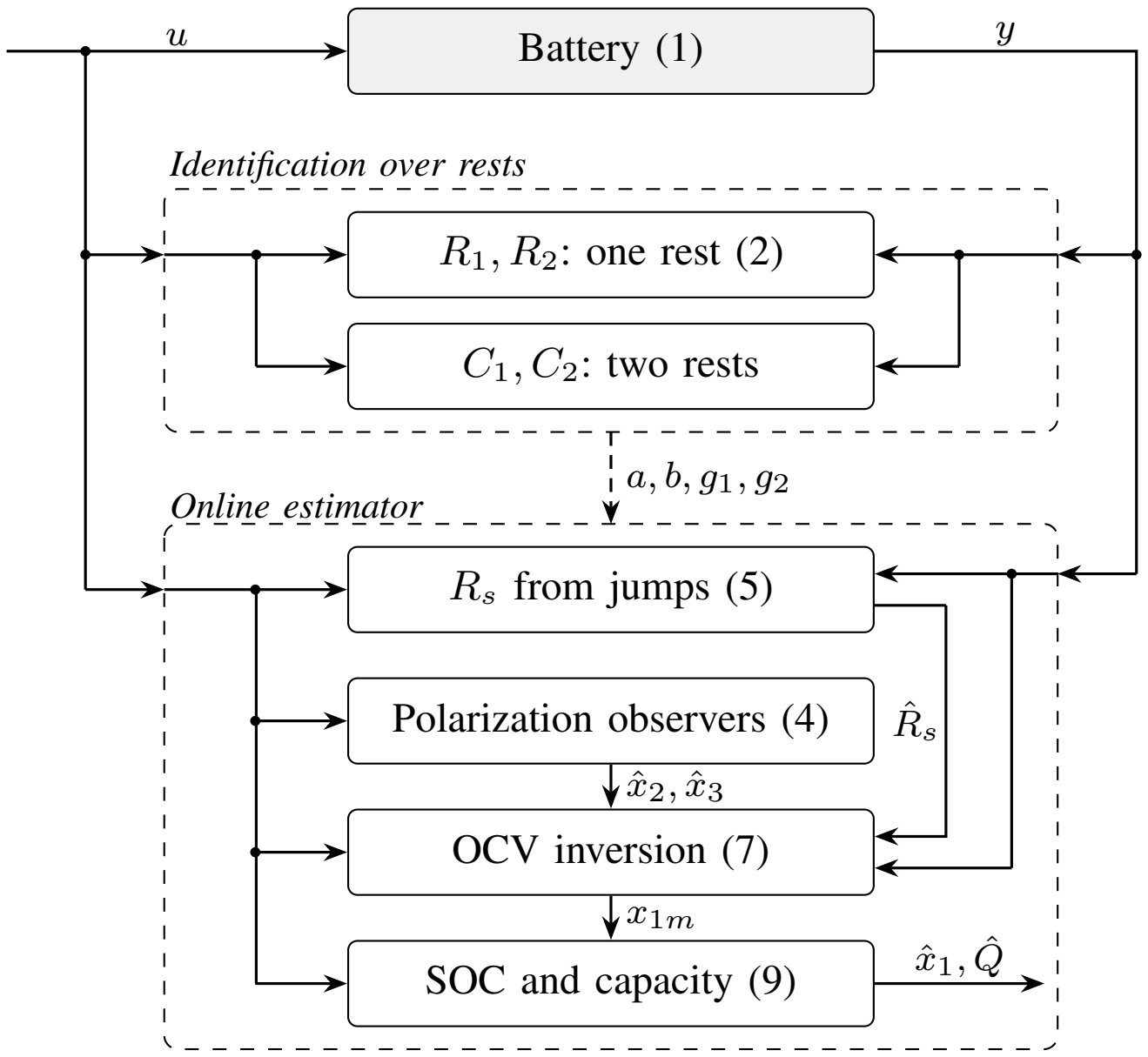
\begin{figure}[ht]
\centering
\resizebox{\columnwidth}{!}{%
\begin{tikzpicture}[
    font=\small,
    blk/.style={draw,rounded corners=2pt,align=center,
        minimum width=40mm,minimum height=6.5mm,inner sep=2pt},
    sig/.style={-{Stealth[length=1.8mm]},semithick},
    lin/.style={semithick},
    lbl/.style={font=\footnotesize,inner sep=1pt},
    grp/.style={draw,dashed,rounded corners=3pt,inner sep=0pt},
    dot/.style={circle,fill,inner sep=0pt,minimum size=2.2pt},
]
\def\xin{2.55}   
\def\xib{3.05}   
\def\xbox{3.4}   
\def\xout{4.0}   

\node[blk,fill=gray!12] (bat) at (0,0) {Battery~\eqref{eq:battery}};

\node[blk] (rid) at (0,-1.55) {$R_1,R_2$: one rest~\eqref{eq:rc_id}};
\node[blk] (cid) at (0,-2.40) {$C_1,C_2$: two rests};
\draw[grp] (-\xbox,-1.05) rectangle (\xbox,-2.90);
\node[lbl,anchor=south west,font=\footnotesize\itshape] at (-\xbox,-1.05)
    {Identification over rests};

\node[blk] (rs)  at (0,-4.10) {$R_s$ from jumps~\eqref{eq:Rs_estimator}};
\node[blk] (pol) at (0,-5.10) {Polarization observers~\eqref{eq:x23_observer}};
\node[blk] (inv) at (0,-6.10) {OCV inversion~\eqref{eq:xm}};
\node[blk] (reg) at (0,-7.10) {SOC and capacity~\eqref{eq:estimates}};
\draw[grp] (-\xbox,-3.60) rectangle (\xbox,-7.60);
\node[lbl,anchor=south west,font=\footnotesize\itshape] at (-\xbox,-3.60)
    {Online estimator};

\draw[sig] (-\xout-0.6,0) -- node[lbl,above] {$u$} (bat.west);
\node[dot] at (-\xout,0) {};
\draw[lin] (-\xout,0) -- (-\xout,-4.10);
\node[dot] at (-\xout,-1.55) {};
\draw[sig] (-\xout,-1.55) -- (-\xbox,-1.55);
\draw[sig] (-\xout,-4.10) -- (-\xbox,-4.10);
\coordinate (uid) at (-2.7,-1.55);
\draw[lin] (-\xbox,-1.55) -- (uid) -- (-2.7,-2.40);
\node[dot] at (uid) {};
\draw[sig] (uid) -- (rid.west);
\draw[sig] (-2.7,-2.40) -- (cid.west);
\coordinate (ues) at (-2.7,-4.10);
\draw[lin] (-\xbox,-4.10) -- (ues) -- (-2.7,-7.10);
\foreach \b in {rs,pol,inv}
    \node[dot] at ($(\b.west)+(-0.7,0)$) {};
\foreach \b in {rs,pol,inv,reg}
    \draw[sig] ($(\b.west)+(-0.7,0)$) -- (\b.west);

\draw[lin] (bat.east) -- node[lbl,above] {$y$} (\xout,0) -- (\xout,-3.98);
\node[dot] at (\xout,-1.55) {};
\draw[sig] (\xout,-1.55) -- (\xbox,-1.55);
\draw[sig] (\xout,-3.98) -- (\xbox,-3.98);
\draw[lin] (\xbox,-1.55) -- (2.65,-1.55) -- (2.65,-2.40);
\node[dot] at (2.65,-1.55) {};
\draw[sig] (2.65,-1.55) -- (rid.east);
\draw[sig] (2.65,-2.40) -- (cid.east);
\draw[lin] (\xbox,-3.98) -- (\xib,-3.98) -- (\xib,-6.22);
\node[dot] at (\xib,-3.98) {};
\draw[sig] (\xib,-3.98) -- ($(rs.east)+(0,0.12)$);
\draw[sig] (\xib,-6.22) -- ($(inv.east)+(0,-0.12)$);

\draw[lin] ($(rs.east)+(0,-0.12)$) -- (\xin,-4.22);
\draw[sig] (\xin,-4.22) -- node[lbl,left] {$\hat R_s$} (\xin,-5.98)
    -- ($(inv.east)+(0,0.12)$);
\draw[sig] (pol) -- node[lbl,right=2pt] {$\hat x_2,\hat x_3$} (inv);
\draw[sig] (inv) -- node[lbl,right=2pt] {$x_{1m}$} (reg);
\draw[sig] (reg.east) -- node[lbl,above] {$\hat x_1,\hat Q$} (\xbox-0.1,-7.10);

\draw[sig,densely dashed] (0,-2.90) -- node[lbl,right=2pt] {$a,b,g_1,g_2$} (0,-3.60);
\end{tikzpicture}%
}
\caption{Estimation scheme. The identification of $R_1$, $R_2$, $C_1$, $C_2$ runs occasionally with data from rests and gives estimates to the online estimator.}
\label{fig:framework}
\end{figure}

Our proposed scheme, summarized in Fig.~\ref{fig:framework}, consists
of two parts operating on different time scales. The
\emph{identification} part uses the voltage relaxation during one rest interval
to identify $R_1$, $R_2$ when $C_1$, $C_2$ are known. If the capacitances are
also unknown, it uses two rests and the intervening current. These parameters
are then treated as known by the \emph{online estimator}, which runs
continuously and estimates $x(t)$, $R_s$, $Q$. The scheme relies on the three mentioned features of the current, all of which can
be checked from the measured $u$ and are naturally present in driving profiles. These features are \emph{rest intervals} ($u=0$) for the identification of
$R_1$, $R_2$, \emph{jumps} of $u$ for the
identification of $R_s$, and \emph{charge throughput} for the
estimation of the SOC and $Q$.
The estimator uses these features to schedule the updates and propagates the
state estimates by the model between updates, yielding a hybrid estimator.

The OCV map $p$ is the only part of the model that we do not estimate. It is a
static characteristic of the cell, which is commonly obtained offline by measuring the
relaxed terminal voltage at several SOC levels, either during a low-current charge or
discharge or after current pulses followed by rests, and by fitting a parametric
function of the SOC to these data, e.g., a polynomial as here, or combinations of
exponential, logarithmic, and sigmoid terms; see~\cite{petzl2013advancement,pattipati2014open} and references therein.
Since it varies only slowly with aging, it can be re-identified offline from time to
time. Our analysis only uses that $p$ is known and strictly increasing on $\X$.

\subsection{Identification Over Rests}
\label{sec:rc_id}

The resistances $R_1$ and $R_2$ are identified from a rest interval
$[s,s+T_r]$, e.g., a stop while driving, with $C_1$ and $C_2$ known. If
$C_1$ and $C_2$ are also unknown, all four parameters are identified
from two rest intervals separated by a loading interval, as stated later in this section.

Let $N\geq4$, $h:=T_r/N$,
$y_k:=y(s+kh)$ for $k=0,\ldots,N$, and
$\Delta y_k:=y_{k+1}-y_k$ for $k=0,\ldots,N-1$. At the endpoints of the rest interval,
take the voltage limits from within the interval. Define
$$
    \Phi:=
    \begin{pmatrix}
        \Delta y_1&-\Delta y_0\\
        \vdots&\vdots\\
        \Delta y_{N-2}&-\Delta y_{N-3}
    \end{pmatrix},
    \qquad
    Y:=
    \begin{pmatrix}
        \Delta y_2\\ \vdots\\ \Delta y_{N-1}
    \end{pmatrix}.
$$
During a rest, since $u=0$, $y$ is the sum of an unknown constant and two decaying
exponentials. The two unknowns of the differenced regression~\eqref{eq:rc_sigma}
require at least two rows in $\Phi$ for $\Phi^\top\Phi$ to be invertible. Thus, this
procedure requires $N-2\geq2$, or at least \emph{five} voltage samples.

\begin{assumption}
\label{ass:rc_id}
The current satisfies $u=0$ on $(s,s+T_r)$, and the matrix
$\Phi^\top\Phi$ is invertible.
\end{assumption}

The invertibility of $\Phi^\top\Phi$ can be checked from the measured
$y$. The estimates are
\begin{subequations}\label{eq:rc_id}
\begin{align}
    &(\hat\sigma_1, \hat\sigma_2)
    :=
    (\Phi^\top\Phi)^{-1}\Phi^\top Y,
    \label{eq:rc_sigma}\\
    &\hat r_1>\hat r_2 \text{ the roots of } r^2-\hat\sigma_1r+\hat\sigma_2,
    \label{eq:rc_roots}\\
    &\hat R_1:=-h/(C_1\ln\hat r_1),
    \quad
    \hat R_2:=-h/(C_2\ln\hat r_2).
    \label{eq:rc_R}
\end{align}
\end{subequations}
\begin{lemma}
\label{prop:rc_id}
Under Assumptions~\ref{ass:operation} and~\ref{ass:rc_id}, the
identification~\eqref{eq:rc_id} is well defined and gives
$(\hat R_1,\hat R_2)=(R_1,R_2)$.
\end{lemma}

\begin{proof}
During rest intervals, $u=0$. By~\eqref{eq:battery}, $x_1$ is constant,
$x_2(t)=x_2(s)e^{-a(t-s)}$, and $x_3(t)=x_3(s)e^{-b(t-s)}$. Let
$r_1:=e^{-ah}$ and $r_2:=e^{-bh}$, so that $1>r_1>r_2>0$ since
$0<a<b$, because $R_1C_1>R_2C_2$ (see Section~\ref{sec:model}). In
particular, $r_1\neq r_2$, and the larger root of~\eqref{eq:rc_roots}
corresponds to the slow branch.
By~\eqref{eq:battery_output}, it follows that
$y_k=p(x_1(s))-x_2(s)r_1^k-x_3(s)r_2^k$, and the differences eliminate
the unknown constant $p(x_1(s))$. Indeed, we have $\Delta y_k=Ar_1^k+Br_2^k$, with
$A:=(1-r_1)x_2(s)$ and $B:=(1-r_2)x_3(s)$. Since $r_1$ and $r_2$ are
the roots of $r^2-\sigma_1r+\sigma_2$ with $\sigma_1:=r_1+r_2$ and
$\sigma_2:=r_1r_2$, we have
$\Delta y_{k+2}=\sigma_1\Delta y_{k+1}-\sigma_2\Delta y_k$ for
$k=0,\ldots,N-3$, i.e., $Y=\Phi(\sigma_1,\sigma_2)$. Since
$\Phi^\top\Phi$ is invertible,~\eqref{eq:rc_sigma} gives
$(\hat\sigma_1,\hat\sigma_2)=(\sigma_1,\sigma_2)$.
Then,~\eqref{eq:rc_roots} gives $(\hat r_1,\hat r_2)=(r_1,r_2)$, and
since $r_1=e^{-h/(R_1C_1)}$ and $r_2=e^{-h/(R_2C_2)}$, from~\eqref{eq:rc_R}
we get $(\hat R_1,\hat R_2)=(R_1,R_2)$.
\end{proof}

\begin{remark}
\label{rem:rc_id}
The identification~\eqref{eq:rc_id} uses only $y$ on the rest intervals
where $u=0$. It
does not require knowing the OCV map $p$, since $x_1$ is constant on the rest
and the differences $\Delta y_k$ eliminate $p(x_1(s))$, nor does it
require $Q$, $R_s$, or the polarization voltages. The matrix $\Phi^\top\Phi$ is invertible if and only if $x_2(s)\neq0$
and $x_3(s)\neq0$, i.e., both branches are excited at the start of
the rest. Indeed, $\Delta y_k=Ar_1^k+Br_2^k$ gives
$\Phi=V\operatorname{diag}(A,B)W$, where $V$ has rows
$(r_1^k,r_2^k)$, $k=0,\ldots,N-3$, and rank two since $r_1\neq r_2$ and
$N\geq4$, and $W$ has rows $(r_1,-1)$ and $(r_2,-1)$ and is invertible. This holds generically after a
loading interval. With measurement noise, the least-squares
fit~\eqref{eq:rc_sigma} uses all samples, and voltage errors enter both $Y$ and
$\Phi$. The exact identification result applies to noise-free data. The
sampling period $h$ should be small compared with the fast time constant
$R_2C_2$, and $T_r$ should be of the order of the slow time constant
$R_1C_1$, so that the slow relaxation is clearly visible.
\end{remark}

If $C_1$ and $C_2$ are also unknown, an earlier rest is added. Let $s_0<s$ and $N_0\geq2$ be such that $u=0$ on $(s_0-N_0h,s_0)$. Define $z_k:=y(s_0-N_0h+kh)$ for $k=0,\ldots,N_0$ and
$\Delta z_k:=z_{k+1}-z_k$ for $k=0,\ldots,N_0-1$, again taking the endpoint
voltage limits from within the rest. With $\hat r_1$, $\hat r_2$
from~\eqref{eq:rc_roots}, let $\hat a:=-\ln(\hat r_1)/h$,
$\hat b:=-\ln(\hat r_2)/h$, and\begin{subequations}\label{eq:c_id}
\begin{align}
    \hat x_2(s)&:=\frac{\Delta y_1-\hat r_2\Delta y_0}{(\hat r_1-\hat r_2)(1-\hat r_1)},
    \hat x_3(s):=\frac{\hat r_1\Delta y_0-\Delta y_1}{(\hat r_1-\hat r_2)(1-\hat r_2)},
    \nonumber\\
    \hat x_2(s_0)&:=\frac{\hat r_1^{N_0}(\Delta z_1-\hat r_2\Delta z_0)}{(\hat r_1-\hat r_2)(1-\hat r_1)},
    \hat x_3(s_0):=\frac{\hat r_2^{N_0}(\hat r_1\Delta z_0-\Delta z_1)}{(\hat r_1-\hat r_2)(1-\hat r_2)},
    \label{eq:c_states}\\
    \hat g_1&:=(\hat x_2(s)-e^{-\hat a(s-s_0)}\hat x_2(s_0))/(\textstyle\int_{s_0}^se^{-\hat a(s-\tau)}u(\tau)d\tau),
    \nonumber\\
    \hat g_2&:=(\hat x_3(s)-e^{-\hat b(s-s_0)}\hat x_3(s_0))/(\textstyle\int_{s_0}^se^{-\hat b(s-\tau)}u(\tau)d\tau),
    \label{eq:c_g}
\end{align}
\end{subequations}
and $\hat C_i:=1/\hat g_i$ ($i=1,2$), $\hat R_1:=\hat g_1/\hat a$, and
$\hat R_2:=\hat g_2/\hat b$. We then state the following lemma.

\begin{lemma}
\label{lem:c_id}
Suppose that Assumptions~\ref{ass:operation} and~\ref{ass:rc_id} hold,
that $u=0$ on $(s_0-N_0h,s_0)$, and that $u$ does not change sign on
$[s_0,s]$ and is not almost everywhere zero there. Then, the identification~\eqref{eq:rc_sigma}-\eqref{eq:rc_roots}-\eqref{eq:c_id} is well
defined and gives $(\hat R_1,\hat R_2,\hat C_1,\hat C_2)=(R_1,R_2,C_1,C_2)$.
\end{lemma}

\begin{remark}
\label{rem:c_id}
The proof of Lemma~\ref{lem:c_id} is similar to that of Lemma~\ref{prop:rc_id} and is not presented here. Indeed, the previous proof gives $\hat a=a$ and
$\hat b=b$ without using either capacitance. The relaxation gives the polarization
voltages in~\eqref{eq:c_states} at the start of the second rest and at the end of the
first, by solving $\Delta y_0=A+B$ and
$\Delta y_1=Ar_1+Br_2$ for $A$ and $B$ as defined in the proof of
Lemma~\ref{prop:rc_id}, and similarly with $\Delta z_0$, $\Delta z_1$. Then,~\eqref{eq:battery_x2}--\eqref{eq:battery_x3} over the loading
interval $[s_0,s]$  give~\eqref{eq:c_g}, where the sign condition
on $u$ ensures that the denominators are non-zero. As for
Lemma~\ref{prop:rc_id}, the OCV map $p$, $Q$, and $R_s$ are not needed, and the first rest may start with both polarization voltages at zero. Its
duration must allow at least three samples ($N_0\geq2$). Identifying $C_1$ and $C_2$
uses the measured $u$ on the loading interval and therefore depends on the accuracy of
that current measurement.
\end{remark}

Since $R_1$, $R_2$, $C_1$, and $C_2$ vary slowly, e.g., with aging and
temperature, this identification does not have to run continuously: it is repeated from
time to time, whenever a suitable rest occurs. From now on, $R_1$, $R_2$, $C_1$, and $C_2$, hence $a$, $b$, $g_1$, and $g_2$, are treated as known.

\subsection{Online Estimator}
\label{sec:online}

With $a$, $b$, $g_1$, $g_2$ known, the online estimator runs
continuously on the measured $u$ and $y$. It combines four steps.
\begin{enumerate}
\item Open-loop observers of $x_2$ and $x_3$, which converge
exponentially since their dynamics are stable, known, and driven by
the measured $u$;
\item An estimator of $R_s$, updated at each jump of $u$, which
exploits the fact that the states are continuous whereas the ohmic
drop $R_su$ jumps with $u$;
\item An OCV inversion, which removes the polarization and ohmic drops
from $y$ and inverts $p$ to obtain a voltage-based SOC signal $x_{1m}$;
\item An estimator of $x_1$ and $\theta$, which propagates $x_1$
by Coulomb counting with the current estimate of $\theta$ and updates both periodically by regressing $x_{1m}$ on the accumulated charge over a sliding window.
\end{enumerate}
The four steps are detailed below.

\subsubsection{Polarization Voltages}
 
With $u$ known, the polarization voltages are estimated by
\begin{equation}\label{eq:x23_observer}
    \dot{\hat x}_2=-a\hat x_2+g_1u,
    \qquad
    \dot{\hat x}_3=-b\hat x_3+g_2u.
\end{equation}
These are copies of~\eqref{eq:battery_x2}--\eqref{eq:battery_x3},
and their estimation errors decay exponentially at rates $a$ and $b$, respectively.
 
\subsubsection{Series Resistance from Current Jumps}
 
Let $t_1<t_2<\cdots$ be the jump times of $u$, which are known
since $u$ is measured, and define the jumps
$$
    \delta u_k := u(t_k^+)-u(t_k^-)\neq0,\qquad
    \delta y_k:=y(t_k^+)-y(t_k^-).
$$
The estimate $\hat R_s$ is held constant between jumps and updated at each jump according to
\begin{subequations}\label{eq:Rs_estimator}
\begin{align}
    t\neq t_k:&\quad
    \dot{\hat R}_s=0,
    \qquad
    \dot S=-\mu S,
    \label{eq:Rs_flow}\\
    t=t_k:&\quad
    S^+=S+\delta u_k^2,
    \nonumber\\
    &\quad
    \hat R_s^+=\hat R_s-(\delta u_k/S^+)
    (\delta y_k+\hat R_s\delta u_k),
    \label{eq:Rs_jump}
\end{align}
\end{subequations}
with $S(0)=0$, an arbitrary constant $\hat R_s(0)$, and a forgetting
factor $\mu\geq0$. Here, $(\cdot)^+$ denotes the value right after the
update. We take the estimator signals to be right-continuous, so $\hat R_s$
keeps its initial value on $[0,t_1)$. The variable
$S(t)=\sum_{t_k\leq t}e^{-\mu(t-t_k)}\delta u_k^2$ is the excitation
accumulated over past jumps. It normalizes the update gain
$\delta u_k/S^+$, so that each new jump is weighted against the
excitation already collected. Indeed, for $t\geq t_1$, $\hat R_s(t)$
minimizes $\sum_{t_k\leq t}e^{-\mu(t-t_k)}(\delta y_k+r\delta u_k)^2$ over
$r\in\R$, i.e., it is a weighted least-squares fit over all past
jumps. For $\mu=0$, all jumps have equal weights in the least-squares
objective. A forgetting factor $\mu>0$
discounts old jumps, which allows $\hat R_s$ to track slow variations
of $R_s$ (e.g., with temperature or aging) at the price of less noise
averaging. The analysis below holds for any $\mu\geq0$.
 
\begin{remark}
\label{rem:udds}
Standard driving profiles such as the UDDS are specified at
1~(Hz), so the resulting current is piecewise constant and can jump at sampling instants. If $u$ has no jumps, the
estimator~\eqref{eq:Rs_estimator} retains its initial value. If $R_s$ is known, it can
be used directly in the voltage correction.
\end{remark}
 
\subsubsection{SOC and Capacity}

Define the \emph{corrected} measurement
\begin{equation}
\overline y:=y+\hat x_2+\hat x_3+\hat R_su,
\label{ycorrected}
\end{equation}
which is known since $u$, $y$, and the estimates $\hat{x}_2$,
$\hat{x}_3$, and $\hat{R}_s$ are known. Define the inversion
\begin{equation}    \label{eq:xm}
    x_{1m}:=p^{-1}(\Pip(\overline y)),
\end{equation}
where $\Pip(v):=\min\{p(1),\max\{p(0),v\}\}$, so that
$x_{1m}\in\X$ for every $\overline y$. If $\hat x_2$, $\hat x_3$, and
$\hat R_s$ were exact, then $\overline y=p(x_1)$ and $x_{1m}=x_1$. Hence, $x_{1m}$ is a
voltage-based measurement of the SOC, whose error vanishes as these estimates converge.
This measurement alone does not give $Q$. To estimate $Q$, we use Coulomb counting,
which relates the SOC at different times through $\theta$ and the charge drawn. Let
$$
q(t):=\int_0^tu(\tau)d\tau
$$
and, for a window length $T>0$ and $\tau\in W_t:=[t-T,t]$,
$$
    \phi_t(\tau):=
    (1, q(t)-q(\tau)),
    \qquad
    \vartheta(t):=
    (x_1(t),\theta).
$$
Integrating~\eqref{eq:battery_x1} from $\tau$ to $t$ gives the linear
regression
\begin{equation}
    x_1(\tau)=\phi_t(\tau)^\top\vartheta(t),
    \qquad
    \tau\in W_t.
    \label{eq:regression}
\end{equation}
Replacing $x_1$ by $x_{1m}$ in~\eqref{eq:regression} gives a linear least-squares
problem in $\vartheta(t)$ over the window $W_t$, whose normal equations involve
$$
M(t):=\frac1T\int_{W_t}\phi_t\phi_t^\top d\tau,\qquad
\beta(t):=\frac1T\int_{W_t}\phi_t x_{1m}d\tau.
$$
When $M(t)$ is invertible, the least-squares solution is $M(t)^{-1}\beta(t)$.
Rather than recomputing it continuously, we use it to reset the estimates periodically,
and propagate them by Coulomb counting in between.
Let $T_u>0$ be an update period and $\tau_j:=\tau_0+jT_u$,
$j\in\N$, the update times, with $\tau_0\geq t_1+T$, so that every regression
window starts at or after the first jump and uses the updated $\hat R_s$. The SOC and
the capacity are estimated by
\begin{subequations}\label{eq:estimates}
\begin{align}
    t\neq\tau_j:&\quad
    \dot{\hat x}_1=-\hat\theta u,
    \quad
    \dot{\hat\theta}=0,
    \label{eq:x1_flow}\\
    t=\tau_j:&\quad
    (\hat x_1^+,\hat\theta^+)
    =
    M(\tau_j)^{-1}\beta(\tau_j),
    \label{eq:x1_jump}
\end{align}
\end{subequations}
with arbitrary initial values $(\hat x_1(0),\hat\theta(0))$, and $\hat Q(t):=1/\hat\theta(t)$ when $\hat\theta(t)\neq0$, with an arbitrary finite value
assigned when $\hat\theta=0$. Between
updates, $\hat x_1$ is propagated by Coulomb counting with the current
estimate of the capacity, and $\hat\theta$ is held constant. At each
update time, both are reset to the solution of the sliding-window
regression. We make the following assumption.

\begin{assumption}
\label{ass:pe}
There exists $\alpha>0$ such that $M(\tau_j)\geq\alpha I$ for all
$j\in\N$.
\end{assumption}

Since $\phi_t$ depends only on $u$, so does $M(t)$, and
Assumption~\ref{ass:pe} can be checked from the measured current. It requires the
accumulated charge to vary within every regression window, uniformly over the windows.
The following lemma gives a simple sufficient condition.

\begin{lemma}
\label{lem:pe}
Suppose that Assumption~\ref{ass:operation} holds and that there exist
constants $\ell\in(0,T]$ and $u_*>0$ such that every
regression window $W_{\tau_j}$ contains an interval
$I_j$ of length $\ell$ on which either $u\geq u_*$ almost everywhere or
$u\leq-u_*$ almost everywhere (cannot change sign).
Then, Assumption~\ref{ass:pe} holds.
\end{lemma}

\begin{proof}
Fix $j\in\N$. By the definition of $\phi_{\tau_j}$ and $M$, we have
$M(\tau_j)=\frac1T\int_{W_{\tau_j}}
\left(\begin{smallmatrix}1& q(\tau_j)-q(\tau)\\ q(\tau_j)-q(\tau)&(q(\tau_j)-q(\tau))^2\end{smallmatrix}\right)d\tau$,
so that $\det M(\tau_j)=\frac1T\int_{W_{\tau_j}}q^2d\tau-\left(\frac1T\int_{W_{\tau_j}}qd\tau\right)^2$,
since the terms in $q(\tau_j)$ cancel. Expanding the square in the following double
integral shows that $
\det M(\tau_j)=
\frac{1}{2T^2}
\int_{W_{\tau_j}}\int_{W_{\tau_j}}
(q(s)-q(r))^2dsdr$.
On $I_j$, $u$ has a constant sign and $|u|\geq u_*$ almost everywhere, so that
$|q(s)-q(r)|\geq u_*|s-r|$ for all $s,r\in I_j$. So, we get
$
\det M(\tau_j)
\geq
\frac{u_*^2}{2T^2}
\int_{I_j}\int_{I_j}(s-r)^2dsdr
=
\frac{u_*^2\ell^4}{12T^2}$.
Since $M(\tau_j)$ is symmetric positive semidefinite of dimension two, its
eigenvalues satisfy $\lmin\lambda_{\max}=\det M(\tau_j)$ and
$\lambda_{\max}\leq\tr M(\tau_j)$. Moreover, $|q(\tau_j)-q(\tau)|\leq\overline uT$ on
$W_{\tau_j}$ gives $\tr M(\tau_j)\leq1+\overline u^2T^2$. Hence, we obtain
$
\lmin(M(\tau_j))
\geq
\frac{\det M(\tau_j)}{\tr M(\tau_j)}
\geq
\frac{u_*^2\ell^4}
{12T^2(1+\overline u^2T^2)}
=:\alpha$,
which recovers Assumption~\ref{ass:pe}.
\end{proof}

\begin{remark}
\label{rem:pe}
The condition of Lemma~\ref{lem:pe} is mild. A constant current $u\equiv u_0\neq0$
satisfies it with $\ell=T$. A piecewise-constant current, e.g., UDDS,
satisfies it with $\ell$ the sampling period of the profile and $u_*$ the smallest
non-zero $|u|$, as soon as $T$ exceeds the longest rest by at least $2\ell$, since
every window contains a full sampling interval on which $u$ is a non-zero
constant. Note that a non-zero current on each window is not sufficient, since the
bound in Assumption~\ref{ass:pe} must be uniform over the windows. Conversely,
Assumption~\ref{ass:pe} fails if $u=0$ almost everywhere on a window, in which case
$\theta$ is not identifiable. An update can also be skipped when $M(\tau_j)$ is poorly
conditioned, e.g., during a long rest, and $\hat x_1$ is then propagated by Coulomb
counting. The analysis below extends to this case if the accepted windows satisfy
Assumption~\ref{ass:pe} and the time between accepted updates is bounded, with this
bound in place of $T_u$.
\end{remark}

\subsection{Exponential Stability of Online Estimator}
\label{sec:analysis}

We now show that, in the nominal case, the online estimator recovers $R_s$ exactly
at the first jump, and that all other estimates converge exponentially. The
polarization observers converge first, which makes $x_{1m}$ converge to $x_1$, and
the periodic updates then transfer this convergence to $\hat x_1$ and $\hat Q$.

\begin{theorem}
\label{thm:nominal}
Consider~\eqref{eq:battery} with $a$, $b$, $g_1$, and $g_2$ known and
the estimator~\eqref{eq:x23_observer},~\eqref{eq:Rs_estimator},~\eqref{eq:xm},~\eqref{eq:estimates}.
Suppose that Assumptions~\ref{ass:operation} and~\ref{ass:pe} hold and
$u$ jumps at least once. Then, $
\hat R_s(t)=R_s$ for all $t\geq t_1$,
and there exist $\overline t\geq\tau_0$, $c>0$, $\lambda>0$ such
that
\begin{equation}
    |(x(t),Q)-(\hat x(t),\hat Q(t))|
    \leq
    ce^{-\lambda t}, \qquad \forall t\geq\overline t,
    \label{eq:thm_bound}
\end{equation}
where $\hat x:=(\hat x_1,\hat x_2,\hat x_3)$.
\end{theorem}

\begin{proof}
\emph{Polarization voltages.} Let $e_2:=x_2-\hat x_2$ and
$e_3:=x_3-\hat x_3$. From~\eqref{eq:battery_x2},~\eqref{eq:battery_x3},
and~\eqref{eq:x23_observer}, $\dot e_2=-ae_2$ and $\dot e_3=-be_3$, so
that
\begin{equation}
    |e_2(t)|+|e_3(t)|\leq E_0e^{-\lambda t},
    \qquad
    t\geq0,
    \label{eq:pol_decay}
\end{equation}
with $\lambda:=\min\{a,b\}$ and $E_0:=|e_2(0)|+|e_3(0)|$.

\emph{Series resistance.} Since $u$ is bounded and piecewise
continuous, $x_1$, $x_2$, and $x_3$ are continuous, and so is $p(x_1)$.
Hence~\eqref{eq:battery_output} gives $\delta y_k=-R_s\delta u_k$ for
every $k$. At $t_1$, $S(t_1^-)=0$, so~\eqref{eq:Rs_jump} gives
$S^+=\delta u_1^2$ and
$\hat R_s^+=\hat R_s-(\delta y_1+\hat R_s\delta u_1)/\delta u_1=R_s$.
Between jumps, $\hat R_s$ is constant and $S>0$. At each later jump,
$\delta y_k+R_s\delta u_k=0$, so $\hat R_s^+=R_s$. Thus, we have
$\hat R_s(t)=R_s$ for $t\geq t_1$.

\emph{SOC signal.} For $t\geq t_1$, $\overline y(t)
=p(x_1(t))-e_2(t)-e_3(t)$. Since $p(x_1)\in[p(0),p(1)]$, we have
$\Pip(p(x_1))=p(x_1)$, and since $|\Pip(v)-\Pip(w)|\leq|v-w|$ for
all $v,w\in\R$, $|\Pip(\overline y)-p(x_1)|\leq|e_2+e_3|$. By
the mean value theorem and $p^\prime\geq m_p$ on $\X$,
$|p^{-1}(v)-p^{-1}(w)|\leq|v-w|/m_p$ for all $v,w\in[p(0),p(1)]$.
Hence, we have $|x_{1m}-x_1|\leq|e_2+e_3|/m_p$, and~\eqref{eq:pol_decay}
gives
\begin{equation}
    |x_1(t)-x_{1m}(t)|
    \leq
    (E_0/m_p)e^{-\lambda t},
    \qquad
    t\geq t_1.
    \label{eq:eta}
\end{equation}

\emph{Updates.} Fix $j\in\N$. Since $\tau_j\geq t_1+T$, we have
$W_{\tau_j}\subset[t_1,\tau_j]$. By~\eqref{eq:regression},
$x_{1m}=\phi_{\tau_j}^\top\vartheta(\tau_j)-(x_1-x_{1m})$ on
$W_{\tau_j}$, and therefore we obtain $
    \vartheta(\tau_j)-M(\tau_j)^{-1}\beta(\tau_j)
    =
    M(\tau_j)^{-1}\frac1T\int_{W_{\tau_j}}\phi_{\tau_j}(x_1-x_{1m})d\tau$.
Since $|q(t)-q(\tau)|\leq\overline uT$ for $\tau\in W_t$, we have
$|\phi_t|\leq\sqrt{1+\overline u^2T^2}$ on $W_t$.
Assumption~\ref{ass:pe} gives $\|M(\tau_j)^{-1}\|\leq1/\alpha$.
Together with~\eqref{eq:eta} and~\eqref{eq:x1_jump}, we get
\begin{equation}
    |(x_1(\tau_j),\theta)-(\hat x_1(\tau_j^+),\hat\theta(\tau_j^+))|
    \leq
    \kappa e^{-\lambda\tau_j},
    \label{eq:vartheta_bound}
\end{equation}
where
$\kappa:=\sqrt{1+\overline u^2T^2}E_0e^{\lambda T}/(\alpha m_p)$.

\emph{Between updates.} Let $t\in[\tau_j,\tau_{j+1})$. By~\eqref{eq:battery_x1}
and~\eqref{eq:x1_flow}, $\hat\theta(t)=\hat\theta(\tau_j^+)$ and
$\frac{d}{d t}(x_1-\hat x_1)=-(\theta-\hat\theta(\tau_j^+))u$, so that $
    |x_1(t)-\hat x_1(t)|
    \leq
    |x_1(\tau_j)-\hat x_1(\tau_j^+)|
    +\overline uT_u|\theta-\hat\theta(\tau_j^+)|$.
Since $e^{-\lambda\tau_j}\leq e^{\lambda T_u}e^{-\lambda t}$,~\eqref{eq:vartheta_bound} gives, for all $t\geq\tau_0$,
\begin{equation}
    |x_1(t)-\hat x_1(t)|+|\theta-\hat\theta(t)|
    \leq
    \kappa^\prime e^{-\lambda t},
    \label{eq:between_bound}
\end{equation}
where $\kappa^\prime :=(2+\overline uT_u)e^{\lambda T_u}\kappa$.

\emph{Capacity.} Let $\overline t\geq\tau_0$ be such that
$2\kappa^\prime Qe^{-\lambda\overline t}\leq1$. For $t\geq\overline t$,~\eqref{eq:between_bound} gives $|\theta-\hat\theta(t)|\leq\theta/2$,
so that $\hat\theta(t)\geq\theta/2>0$ and we have $
    |Q-\hat Q(t)|
    =
    \frac{|\theta-\hat\theta(t)|}{\theta\hat\theta(t)}
    \leq
    2Q^2|\theta-\hat\theta(t)|
    \leq
    2Q^2\kappa^\prime e^{-\lambda t}$.
Combining with~\eqref{eq:pol_decay} and~\eqref{eq:between_bound},
and using that the Euclidean norm is bounded by the sum of the
absolute values of the components, we get~\eqref{eq:thm_bound} with
$c:=E_0+(1+2Q^2)\kappa^\prime$.
\end{proof}

\subsection{Robustness of Online Estimator}
\label{sec:robust}

In practice, the current and voltage measurements are corrupted, and the model is
not exact. We now consider additive disturbances in the battery dynamics and
measurement noise:\begin{subequations}\label{eq:battery_perturbed}
\begin{align}
    \dot x_1 &= -\theta u+d_1,
    \label{eq:pert_x1}\\
    \dot x_2 &= -a x_2+g_1u+d_2,
    \label{eq:pert_x2}\\
    \dot x_3 &= -b x_3+g_2u+d_3,
    \label{eq:pert_x3}\\
    y &= p(x_1)-x_2-x_3-R_su+w,
    \label{eq:pert_y}
\end{align}
\end{subequations}
where $d_1,d_2,d_3$ denote process disturbances and $w$ denotes
measurement noise. The online estimator is kept unchanged, and
$a$, $b$, $g_1$, and $g_2$ are assumed to be known. For example, $d_1$ can represent
current-sensor bias in Coulomb counting, while $d_2$ and $d_3$ can represent
current-sensor errors and unmodeled polarization dynamics. Note that voltage-sensor noise $w$
may jump at the jump times of $u$ (see Section~\ref{subsec:udds}). It can also model, e.g., inexact identification of $p$.

\begin{assumption}
\label{ass:bound}
The signals $d_1,d_2,d_3,w$ are piecewise continuous, and there exist
$\overline d_1,\overline d_2,\overline d_3,\overline w\geq0$ such that
\begin{equation}
    |d_i(t)|\leq\overline d_i,
    \qquad
    |w(t)|\leq\overline w,
    \qquad
    \forall t\geq0.
    \label{eq:disturbance_bounds}
\end{equation}
Moreover, the SOC satisfies $x_1(t)\in\X$ for all $t\geq0$.
\end{assumption}

The noise $w$ may jump at any time, including at the jump times $t_k$
of the current. Its effect on $\hat R_s$ is controlled by the
weighted root-mean-square size of the current jumps,
$$
    \rho(t)
    :=
    \left(
    \frac{\sum_{t_k\leq t}e^{-\mu(t-t_k)}\delta u_k^2}
         {\sum_{t_k\leq t}e^{-\mu(t-t_k)}}
    \right)^{1/2},
    \qquad t\geq t_1.
$$

\begin{assumption}
\label{ass:jumps}
There exists $\rho_0>0$ such that $\rho(t)\geq\rho_0$ for all
$t\geq t_1$.
\end{assumption}

\begin{remark}
Assumption~\ref{ass:jumps} holds, for instance, if all jumps satisfy
$|\delta u_k|\geq\rho_0$, which can be enforced by ignoring current jumps below a threshold. If so, $t_k$ denotes the retained
jump times in the $R_s$ estimator and its analysis, and the regression windows
start at or after the first retained jump.
\end{remark}

\begin{theorem}
\label{thm:robust}
Consider~\eqref{eq:battery_perturbed} with $a$, $b$, $g_1$, $g_2$ known
and the
estimator~\eqref{eq:x23_observer},~\eqref{eq:Rs_estimator},~\eqref{eq:xm},~\eqref{eq:estimates}.
Suppose that
Assumptions~\ref{ass:operation},~\ref{ass:pe},~\ref{ass:bound},
and~\ref{ass:jumps} hold and $u$ jumps at least once. Then, we have $
|R_s-\hat R_s(t)|\leq2\overline w/\rho_0$ for all $t\geq t_1$.
There exist $c,\lambda,\gamma,r>0$ such that, if
$\overline d:=\overline d_1+\overline d_2+\overline d_3+\overline w\leq r$,
then there exists $\overline t\geq\tau_0$ such that
\begin{equation}
    |(x(t),Q)-(\hat x(t),\hat Q(t))|
    \leq
    ce^{-\lambda t}+\gamma\overline d, \qquad \forall t\geq\overline t.
    \label{eq:thm_robust}
\end{equation}
\end{theorem}

\begin{proof}
\emph{Polarization voltages.} With $e_2:=x_2-\hat x_2$ and
$e_3:=x_3-\hat x_3$,~\eqref{eq:pert_x2},~\eqref{eq:pert_x3},
and~\eqref{eq:x23_observer} give $\dot e_2=-ae_2+d_2$ and
$\dot e_3=-be_3+d_3$. From
Assumption~\ref{ass:bound}, we get
\begin{equation}
    |e_2(t)|+|e_3(t)|\leq E_0e^{-\lambda t}+D_p,
    \qquad
    t\geq0,
    \label{eq:robust_pol}
\end{equation}
with $\lambda$, $E_0$ in the proof of Theorem~\ref{thm:nominal} and
$D_p:=\overline d_2/a+\overline d_3/b$.

\emph{Series resistance.} Since the right-hand side
of~\eqref{eq:battery_perturbed} is bounded, the states are continuous,
whereas $w$ may jump. Hence~\eqref{eq:pert_y} gives
$\delta y_k=-R_s\delta u_k+\delta w_k$, with
$\delta w_k:=w(t_k^+)-w(t_k^-)$ and $|\delta w_k|\leq2\overline w$. For
$t\geq t_1$, $\hat R_s(t)$ is the weighted least-squares fit of
$\delta y_k=-r\delta u_k$ with weights $\omega_k:=e^{-\mu(t-t_k)}$, so
that $
    R_s-\hat R_s(t)
    =
    \frac{\sum_{t_k\leq t}\omega_k\delta u_k\delta w_k}
         {\sum_{t_k\leq t}\omega_k\delta u_k^2}$.
By the Cauchy--Schwarz inequality and Assumption~\ref{ass:jumps}, we get $
    |R_s-\hat R_s(t)|
    \leq
    \left(\frac{\sum_{t_k\leq t}\omega_k\delta w_k^2}
               {\sum_{t_k\leq t}\omega_k\delta u_k^2}\right)^{1/2}
    \leq
    \frac{2\overline w}{\rho(t)}
    \leq
    \frac{2\overline w}{\rho_0}$.

\emph{SOC signal.} For $t\geq t_1$,~\eqref{ycorrected}
and~\eqref{eq:pert_y} give
$\overline y-p(x_1)=-e_2-e_3+(\hat R_s-R_s)u+w$, so that,
by~\eqref{eq:robust_pol} and the bound on $\hat R_s$,
$|\overline y-p(x_1)|\leq E_0e^{-\lambda t}+D_y$ with
$D_y:=D_p+(1+2\overline u/\rho_0)\overline w$. As in the proof of
Theorem~\ref{thm:nominal}, this gives
\begin{equation}
    |x_1(t)-x_{1m}(t)|
    \leq
    (1/m_p)(E_0e^{-\lambda t}+D_y),
    \quad
    t\geq t_1.
    \label{eq:robust_eta}
\end{equation}

\emph{Updates.} Integrating~\eqref{eq:pert_x1} from $\tau$ to $t$
gives $x_1(\tau)=\phi_t(\tau)^\top\vartheta(t)-\int_\tau^td_1(s)d s$.
Fix $j\in\N$, so that $W_{\tau_j}\subset[t_1,\tau_j]$. Then, we have $\vartheta(\tau_j)-M(\tau_j)^{-1}\beta(\tau_j)   =
    M(\tau_j)^{-1}\frac1T\int_{W_{\tau_j}}\phi_{\tau_j}
    \left(x_1-x_{1m}+\int_\tau^{\tau_j}d_1(s)d s\right)d\tau$,
where $|\int_\tau^{\tau_j}d_1(s)d s|\leq T\overline d_1$ for
$\tau\in W_{\tau_j}$. As in the proof of Theorem~\ref{thm:nominal},
$|\phi_{\tau_j}|\leq\sqrt{1+\overline u^2T^2}$ on $W_{\tau_j}$ and
$\|M(\tau_j)^{-1}\|\leq1/\alpha$. Using~\eqref{eq:robust_eta}
and~\eqref{eq:x1_jump}, we get
\begin{equation}
    |(x_1(\tau_j),\theta)-(\hat x_1(\tau_j^+),\hat\theta(\tau_j^+))|
    \leq
    Ae^{-\lambda\tau_j}+B,
    \label{eq:robust_vartheta}
\end{equation}
where $\kappa_0:=\sqrt{1+\overline u^2T^2}/\alpha$,
$A:=\kappa_0E_0e^{\lambda T}/m_p$, and
$B:=\kappa_0(D_y/m_p+T\overline d_1)$.

\emph{Between updates.} Let $t\in[\tau_j,\tau_{j+1})$.
By~\eqref{eq:pert_x1} and~\eqref{eq:x1_flow},
$\hat\theta(t)=\hat\theta(\tau_j^+)$ and
$\frac{d}{d t}(x_1-\hat x_1)=-(\theta-\hat\theta(\tau_j^+))u+d_1$, so
that we obtain $|x_1(t)-\hat x_1(t)|
    \leq
    |x_1(\tau_j)-\hat x_1(\tau_j^+)|
    +\overline uT_u|\theta-\hat\theta(\tau_j^+)|
    +T_u\overline d_1$.
Since $e^{-\lambda\tau_j}\leq e^{\lambda T_u}e^{-\lambda t}$,~\eqref{eq:robust_vartheta} gives, for all $t\geq\tau_0$,
\begin{equation}
    |x_1(t)-\hat x_1(t)|+|\theta-\hat\theta(t)|
    \leq
    A^\prime e^{-\lambda t}+B^\prime,
    \label{eq:robust_between}
\end{equation}
where $A^\prime :=(2+\overline uT_u)e^{\lambda T_u}A$ and
$B^\prime :=(2+\overline uT_u)B+T_u\overline d_1$. Note that
$B^\prime \leq K\overline d$ with $K:=(2+\overline uT_u)\kappa_0\max\left\{
    T,\frac{1}{am_p},\frac{1}{bm_p},
    \frac{1}{m_p}\left(1+\frac{2\overline u}{\rho_0}\right)
    \right\}+T_u$.

\emph{Capacity.} Let $r:=1/(4KQ)$, so that $\overline d\leq r$ implies
$B^\prime \leq\theta/4$, and let $\overline t\geq\tau_0$ be such that
$4A^\prime Qe^{-\lambda\overline t}\leq1$. For $t\geq\overline t$,~\eqref{eq:robust_between} gives $|\theta-\hat\theta(t)|\leq\theta/2$,
so $\hat\theta(t)\geq\theta/2>0$ and
$
    |Q-\hat Q(t)|
    = \frac{|\theta-\hat\theta(t)|}{\theta\hat\theta(t)}
    \leq 2Q^2|\theta-\hat\theta(t)| \leq 2Q^2(A^\prime e^{-\lambda t}+B^\prime)$.
Combining with~\eqref{eq:robust_pol} and~\eqref{eq:robust_between},
and using that the Euclidean norm is bounded by the sum of the
absolute values of the components, we get~\eqref{eq:thm_robust} with
$c:=E_0+(1+2Q^2)A^\prime$ and $\gamma:=\max\{1/a,1/b\}+(1+2Q^2)K$, since
$D_p\leq\max\{1/a,1/b\}\overline d$ and $B^\prime \leq K\overline d$.
\end{proof}

\begin{remark}
\label{rem:robust}
For $\overline d=0$, Theorem~\ref{thm:robust} recovers
Theorem~\ref{thm:nominal}. If $w$ is continuous at every jump time
$t_k$, then $\delta w_k=0$, so that $\hat R_s(t)=R_s$ for all $t\geq t_1$,
Assumption~\ref{ass:jumps} is not needed, and the factor
$1+2\overline u/\rho_0$ in $D_y$ and $K$ is replaced by $1$. Between
updates, the drift in the SOC estimation error due to $d_1$ is at most
$T_u\overline d_1$, so that the update period $T_u$ trades off this drift
of Coulomb counting against the frequency of the
regression. The condition
$\overline d\leq r$ is only needed for $\hat Q=1/\hat\theta$.
It can be removed by imposing a lower bound of $1/Q_{\max}$ on
$\hat\theta$ when computing $\hat Q$, where $Q_{\max}$ is a known upper
bound on $Q$. The bound on $\hat R_s$ is a worst-case bound. For zero-mean
noise and $\mu=0$, the least-squares fit averages over the jumps, so
that the error can be smaller in practice.
\end{remark}

\section{ILLUSTRATIONS}
\label{sec:illu}

The MATLAB code that reproduces our estimation scheme and the simulations below, from the UDDS profile of~\cite{EPA_UDDS}, is available at~\cite{code}.
We simulate~\eqref{eq:battery} with the parameters of
Section~\ref{sec:model}, driven by the UDDS current shown in
Fig.~\ref{fig:udds_current}.
The initial state is taken as $x(0)=(0.9,0.02,0.01)$, i.e., the
initial polarization voltages are non-zero. The voltage is sampled
with period $h=0.1$~(s), and only rests lasting at least
$2$~(s), i.e., $N\geq20$, are used for the identification.

\subsection{Identification Over Rests}

\begin{figure}[!t]
    \centering
    \includegraphics[width=\columnwidth]{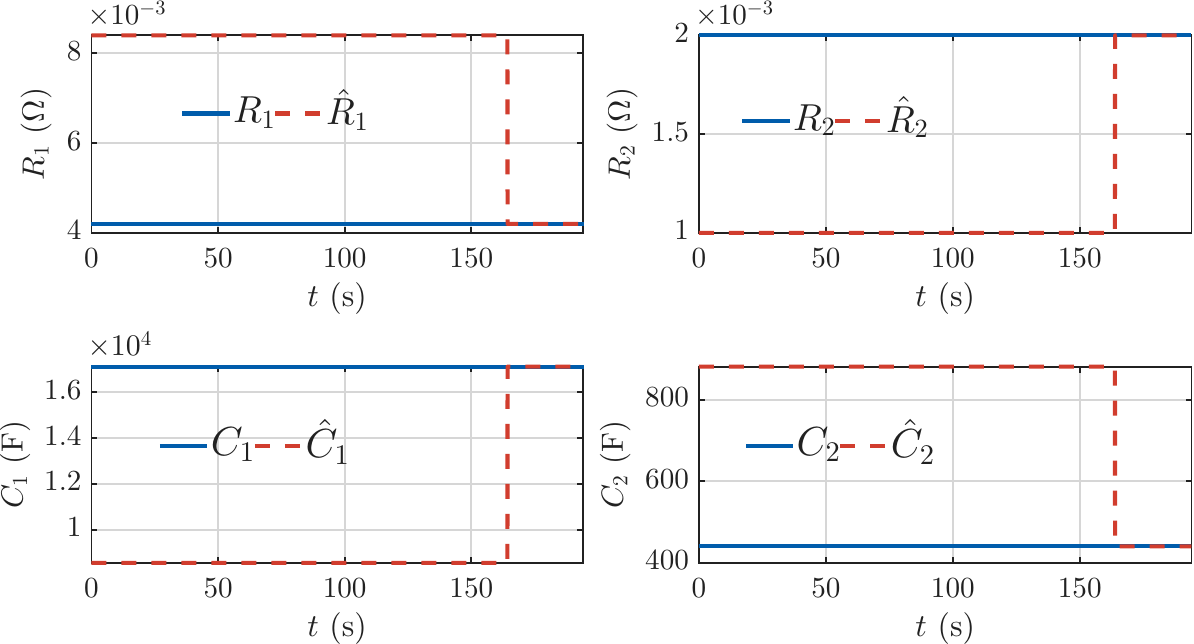}
    \caption{Identification of $R_1$, $R_2$, $C_1$, and $C_2$ over the first two
    rests. The estimates are
    initial guesses until the end of the second rest.}
    \label{fig:id}
\end{figure}

We first illustrate the identification of $R_1$, $R_2$, $C_1$,
and $C_2$ on a short simulation that stops shortly after the second
rest of the profile, at $t=194$~(s). The relaxation during the second
rest gives $a$ and $b$. The endpoint polarization voltages recovered from the two
rests and the intervening current give $g_1$ and $g_2$, as in
Lemma~\ref{lem:c_id}. Before the second rest, the parameters are set to
the initial guesses $2R_1$, $R_2/2$, $C_1/2$, and $2C_2$, e.g., from a
datasheet. Fig.~\ref{fig:id} shows that at the end of the second rest,
at $t=163.9$~(s), all four estimates have near-zero errors, although $p$, $Q$, and $R_s$ are not used.

\subsection{Online Estimator}

We then run the online estimator over the UDDS  profile with $R_1$, $R_2$,
$C_1$, and $C_2$ known, $T=300$~(s), $T_u=60$~(s), $\tau_0=t_1+T+h$, and
$\mu=0$, from $\hat x(0)=(0.5,0,0)$, $\hat R_s(0)=2R_s$, and
$\hat Q(0)=1.3Q$. An update is skipped when $M(\tau_j)$ is poorly
conditioned, as discussed in Remark~\ref{rem:pe}. The UDDS current satisfies the condition of Lemma~\ref{lem:pe} with $\ell=1$~(s),
since its longest rest is shorter than $T-2\ell=298$~(s).

\subsubsection{Nominal Case}

\begin{figure}[!t]
    \centering
    \includegraphics[width=\columnwidth]{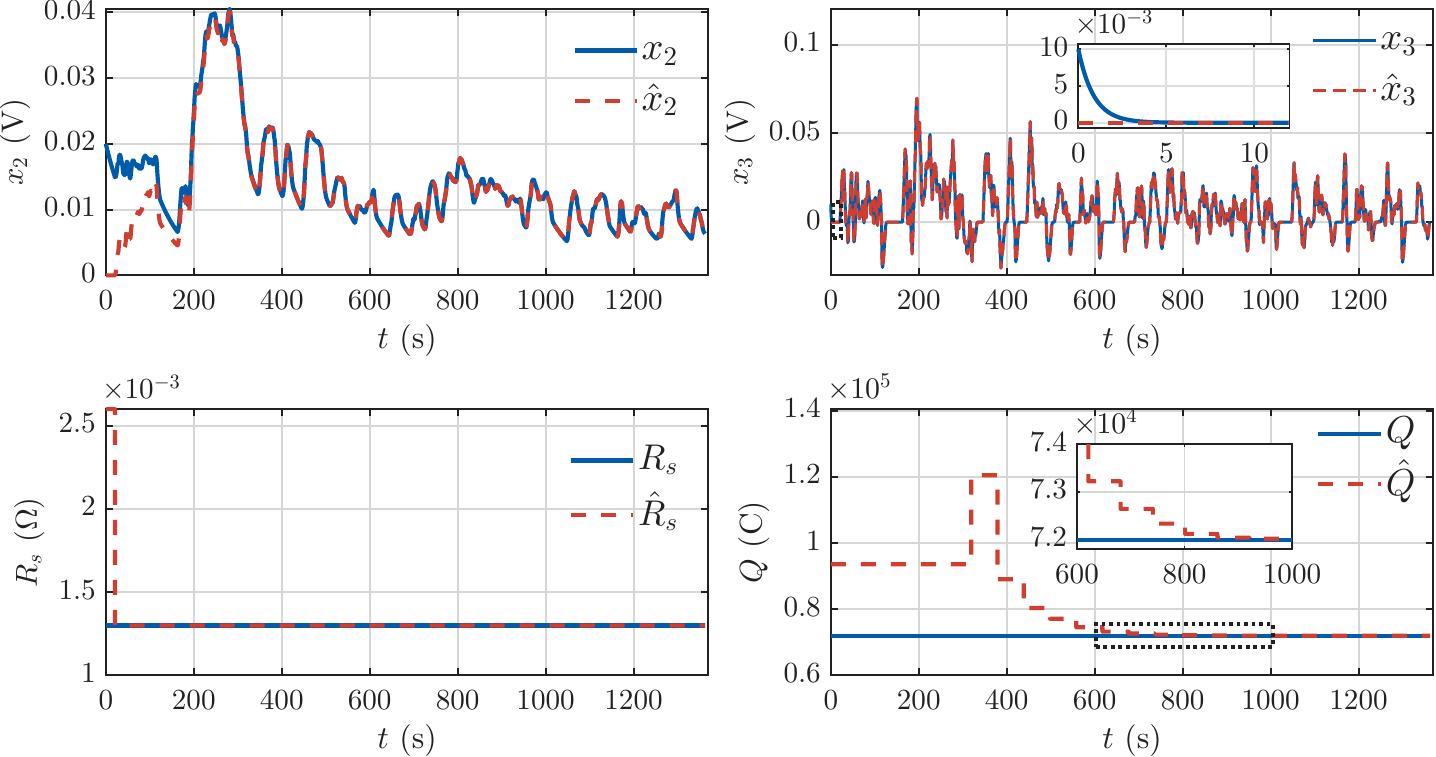}
    \includegraphics[width=\columnwidth]{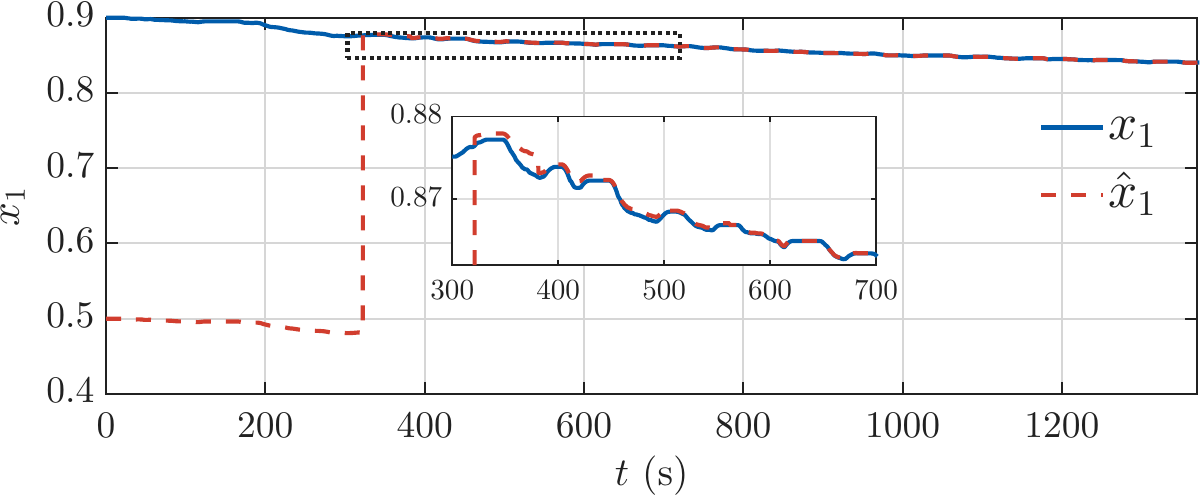}
    \caption{Online estimator in nominal case: $x_2$, $x_3$, $R_s$,
    $Q$ (top); $x_1$ (bottom).}
    \label{fig:nominal}
\end{figure}

Fig.~\ref{fig:nominal} shows the simulation results. The estimate $\hat R_s$
equals $R_s$ from the first jump at $t_1=21$~(s), as stated
in Theorem~\ref{thm:nominal}. The estimate $\hat x_3$ converges within a few
seconds, while $\hat x_2$ converges with the time constant $R_1C_1\approx72$~(s),
since the observers~\eqref{eq:x23_observer} are open-loop copies of the
RC dynamics. The first update of $\hat x_1$ and $\hat Q$ occurs at
$\tau_0=321.1$~(s). The first updates are inaccurate, since their windows still contain the transient of $\hat x_2$ in $\overline y$, hence in $x_{1m}$. The subsequent estimates
improve as that transient decays, as predicted by
Theorem~\ref{thm:nominal}. Between updates, $\hat x_1$ is propagated by Coulomb
counting using the latest capacity estimate.

\subsubsection{With Uncertainties}

\begin{figure}[!t]
    \centering
    \includegraphics[width=\columnwidth]{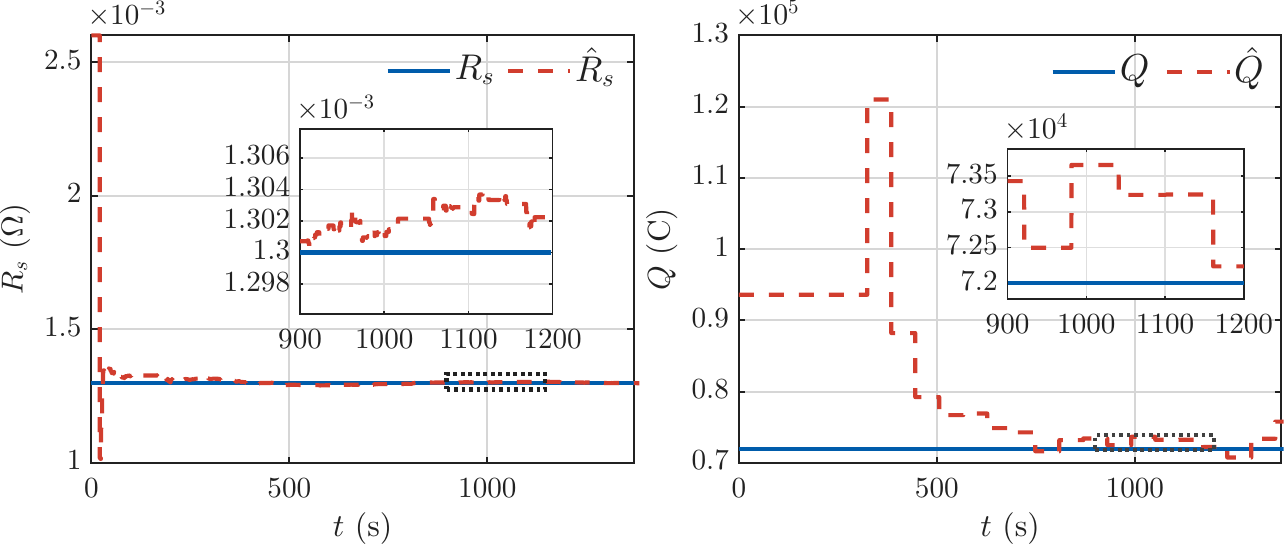}
    \caption{Online estimator with disturbances and noise: $R_s$ (left); $Q$ (right).}
    \label{fig:uncertain}
\end{figure}

We add $d_1=5\times10^{-7}$~(s$^{-1}$), i.e., a current-sensor bias of
about $0.04$~(A), $d_2$ and $d_3$ uniformly distributed in
$[-10^{-4},10^{-4}]$~(V/s), and voltage noise $w$ drawn independently at each
sample from the uniform distribution on $[-1,1]$~(mV). The voltage noise can jump at
the current jump times. Only jumps with
$|\delta u_k|\geq2$~(A) are used, so that Assumption~\ref{ass:jumps}
holds with $\rho_0=2$~(A), and the worst-case bound of
Theorem~\ref{thm:robust} gives $|R_s-\hat R_s|\leq1$~(m$\Omega$).
Fig.~\ref{fig:uncertain} shows $R_s$ and $Q$, along with their estimates, which
show the largest deviations from the nominal case. The state estimates are visually
indistinguishable from those in Fig.~\ref{fig:nominal}. 

As the least-squares fit accumulates data from the retained
jumps, we see that the error in $\hat R_s$ settles below
$0.004$~(m$\Omega$), far below the worst-case
bound. For $t\geq900$~(s), the errors in $x_1$
and $Q$ remain below $0.02$ percentage points
and $5.3\%$, respectively. Each regression update
corrects the SOC estimate after it has drifted
due to the bias $d_1$.

\section{CONCLUSION}
\label{sec:conc}

We proposed a hybrid state-parameter estimation scheme for a second-order
Th\'evenin battery model.
The approach strategically exploits different features of the current for a full state-parameter estimation, with each estimate updated when the
corresponding information becomes available. We established
exact identification results, exponential convergence in the
nominal case, and robustness under uncertainties.

Future work will study the temperature dependence of the
OCV--SOC relation and validate our scheme
experimentally.

\bibliographystyle{IEEEtran}
\bibliography{ref}

\end{document}